\documentclass[runningheads]{llncs}

\usepackage[T1]{fontenc}

\usepackage[utf8]{inputenc}
\usepackage{CJKutf8}

\title{Finite Functional Programming}
\subtitle{or, LAMBDA: The Ultimate Predicate}

\author{Michael Arntzenius\orcidID{0009-0002-0417-5636}
  \and Max Willsey\orcidID{0000-0001-8066-4218}
}
\authorrunning{Arntzenius and Willsey}
\institute{University of California, Berkeley, USA\\
\email{daekharel@gmail.com} \qquad \email{mwillsey@berkeley.edu}
}

\usepackage{color}
\usepackage[dvipsnames,svgnames]{xcolor}

\usepackage{amssymb}            
\usepackage{amsmath}
\usepackage{stmaryrd}           
\usepackage{mathtools}          
\usepackage{mathpartir}         
\usepackage{nccmath}            
\usepackage{booktabs}           
\usepackage{pifont}             
\usepackage{placeins}           
\usepackage{changepage}         
\usepackage{etoolbox}           
\usepackage{graphicx}           

\usepackage{listings}           
\usepackage{url}
\usepackage{hyperref}

\usepackage[nameinlink,noabbrev]{cleveref}

\newlength{\bigwidth}
\newlength{\bigwidthoffset}
\newenvironment{fullwidth}
  {\begin{adjustwidth}{\the\bigwidthoffset}{\the\bigwidthoffset}}
  {\end{adjustwidth}}

\makeatletter
\renewenvironment{figure*}[1][\fps@figure]
  {\edef\@tempa{\noexpand\@float{figure}[#1]}%
   \@tempa\begin{fullwidth}\centering}
  {\end{fullwidth}\end@float}
\makeatother

\usepackage{environ}            
\newlength\codeoffset
\NewEnviron{code}{
  \begin{fleqn}[\codeoffset]
    \begin{flalign*}
      \setlength\arraycolsep{0pt} 
      \begin{array}{l}\BODY\end{array}
  \end{flalign*}
  \end{fleqn}
}

\makeatletter
\newlength{\negph@wd}
\DeclareRobustCommand{\negphantom}[1]{%
  \ifmmode
    \mathpalette\negph@math{#1}%
  \else
    \negph@do{#1}%
  \fi
}
\newcommand{\negph@math}[2]{\negph@do{$\m@th#1#2$}}
\newcommand{\negph@do}[1]{%
  \settowidth{\negph@wd}{#1}%
  \hspace*{-\negph@wd}%
}\makeatother

\newcommand\strong[1]{{\bfseries#1}}
\newcommand\todo[1]{{\color{IndianRed}#1}}
\newcommand\TODO[1]{\todo{\strong{TODO}: #1}}

\newcommand\todolater\todo
\renewcommand\todolater[1]{\todo{\color{RoyalBlue}#1}}

\renewcommand\todo[1]{\ignorespaces}

\newcommand\fslang{\ensuremath{\lambda_{\mathrm{FS}}}}
\newcommand\catname[1]{\ensuremath{\mathrm{#1}}}
\newcommand\Set{\ensuremath{\catname{Set}}}
\newcommand\Point{\ensuremath{\catname{Set}_{*}}}
\DeclareMathOperator{\support}{supp}

\newcommand\den[1]{\llbracket{#1}\rrbracket}
\newcommand\bnfeq{\Coloneqq}
\newcommand\bnfor{\mathrel{\,|\,}}

\newcommand\name[1]{\ensuremath{\mathit{#1}}}
\newcommand\<\:                 
\newcommand\fnspace{\<}         
\newcommand\fn[1]{\lambda{#1}.\fnspace}
\newcommand\fa[1]{\forall{#1}.\fnspace}
\newcommand\ex[1]{\exists{#1}.\fnspace}

\newcommand\cto\to
\newcommand\lto\multimap        
\newcommand\fto\Rightarrow
\newcommand\x\times
\newcommand\with{\mathbin{\&}}
\newcommand\ox\otimes
\newcommand\typename\textit
\newcommand\bool{\typename{bool}}
\newcommand\maybe{\typename{maybe}}
\newcommand\maybeof[1]{\ensuremath{\maybe~{#1}}}

\newcommand\G{\mathrm{\Gamma}}
\newcommand\D{\mathrm{\Delta}}
\renewcommand\O{\mathrm{\Omega}}
\newcommand\emptycx{{\cdot}}
\newcommand\of{\mathbin{:}}
\newcommand\cxsep{\mathbin{/}}
\newcommand\hyp[2]{{#1} \of {#2}}

\newcommand\entails{\,\vdash\,}
\newcommand\cJ[3]{{#3} \entails {#1} : {#2}}
\newcommand\J[5]{{#3} \cxsep {#4} \cxsep {#5} \entails {#1} : {#2}}
\newcommand\lJ[4]{{#3} \cxsep {#4} \entails {#1} : {#2}}
\newcommand\rulename\textsc

\newcommand\kword[1]{\ensuremath{\textrm{#1}}}
\newcommand\const\name
\newcommand\cname\textrm
\newcommand\cnil{\cname{none}}
\newcommand\cjust{\cname{just}}

\newcommand\nil{\name{nil}}
\newcommand\true{\name{true}}
\newcommand\false{\name{false}}

\newcommand\letin[1]{\kword{let}~{#1}~\kword{in}~}
\newcommand\leteq[2]{\letin{{#1} = {#2}}}
\newcommand\tjust[1]{\cjust\<{#1}}
\newcommand\letjust[3]{\leteq{\tjust {#1}}{#2}{#3}}
\newcommand\letpair[4]{\leteq{({#1}, {#2})}{#3}{#4}}

\newcommand\tand{\mathrel{\kword{and}}}
\newcommand\tor{\mathrel{\kword{or}}}
\newcommand\whenkword{\kword{when}}
\newcommand\when{\mathrel{\whenkword}}
\newcommand\teq[2]{{#1} = {#2}}
\newcommand\withpair[1]{\ensuremath{\langle{#1}\rangle}}
\newcommand\ifjust[4]{%
  \kword{case}~{#1}\ \kword{of}\
  \cjust~{#2} \shortrightarrow {#3};\, \cnil \shortrightarrow {#4}%
}

\newcommand\pwild{\text{\textunderscore}}

\newcommand\typeforget{}
\newcommand\termforget[1]{#1}

\newcommand\nat{\mathbb{N}}
\newcommand\natz{\ensuremath{\nat_0}}

\newtoggle{charter}
\togglefalse{charter}

\toggletrue{charter}

\usepackage[scaled=0.965, scaled=0.98, proportional,semibold]{sourcesanspro}

\usepackage[scaled=1.021,var0,varl,]{inconsolata}

\usepackage[scaled=0.97]{XCharter}
\usepackage[xcharter, alty, scaled=1.0556, 
]{newtxmath}

\renewcommand\G\Gamma
\renewcommand\D\Delta
\renewcommand\O\Omega

\begin{document}
\maketitle              

\begin{abstract}
We unify functional and logic 
programming by treating predicates 
as functions equipped with their support: the set of inputs whose output is nonzero.
Datalog, for instance, is a language of finitely supported boolean functions.
Finite support allows representing functions as input-output tables.
Generalizing from boolean functions to other pointed sets neatly handles aggregation and weighted logic programming.
We refer to the combination of finitely supported functions, represented as data, with higher order functions, represented as code, as \emph{finite functional programming.}
We give a simple type system to check finite support, using graded effects to check variable grounding and relevance types to model pointed sets.


\end{abstract}


\section{Logic programming is (not) functional programming}

Consider the following simple logic program to find mutual follows in a social network graph:

\newcommand\pred\lstinline

\begin{lstlisting}
follows(john, mary).
follows(mary, john).
% ... many more follows() facts.
mutuals(X,Y) :- follows(X,Y), follows(Y,X).
\end{lstlisting}



\noindent
Predicates like \pred{follows} and \pred{mutuals} denote functions into booleans.
Following this intuition, we can translate this logic program directly into a functional one:

\begin{code}
\name{follows},\, \name{mutuals} : \name{user} \to \name{user} \to \bool\\
\name{follows} \<x \<y = (\const{john} == x \tand \const{mary} == y)
\\
\phantom{\name{follows} \<x \<y =}
\negphantom{{}\tor}
\tor (\const{mary} == x \tand \const{john} == y)
\\
\phantom{\name{follows} \<x \<y =}
\negphantom{{}\tor}
\tor \ldots
\\
\name{mutuals} \<x \<y = \name{follows} \<x \<y \tand \name{follows} \<y \<x
\end{code}




\noindent
However, functions in functional and imperative languages are unidirectional: they take inputs to outputs.
Evaluating $\name{mutuals} \<x \<y$ will test whether fixed users $x, y$ are mutuals---but unlike in logic programming, we cannot use it to enumerate mutual follows, for instance, to find the mutuals $y$ of a fixed user $x$.%


For the same reason, our functional translation cannot handle existential quantifiers (also called `projection' in databases). Consider this logic program to find actors who have appeared in the same film:

\begin{lstlisting}
costars(Actor1, Actor2) :-
  stars(Film, Actor1),
  stars(Film, Actor2).
\end{lstlisting}


\noindent
Since \lstinline{Film} does not appear in the head of the rule, the head is derivable if there exists a \lstinline{Film} satisfying the body.
To embed this functionally, we could assume a function $\name{exists} : \fa{\alpha} (\alpha \to \bool) \to \bool$:

\begin{code}
\name{costars} : \name{person} \to \name{person} \to \bool\\



\name{costars} \<\name{actor1} \<\name{actor2} =
\name{exists}\<(\fn{\name{film}}
\name{stars} \<\name{film} \<\name{actor1}
\tand
\name{stars} \<\name{film} \<\name{actor2})
\end{code}

\noindent
But how can we implement \name{exists} for an arbitrary type $\alpha$ and function $f : \alpha \to \bool$?
We are stymied by unidirectionality: all we can do with $f$ is give it inputs; since the input type is arbitrary, we cannot generate any.%
\footnote{%
  Of course, one can define \name{exists} for any finite type $\alpha$ by enumeration, and
  even for some infinite types by exploiting continuity of $f$~\cite{seemingly-impossible-functional-programs,DBLP:conf/lics/Escardo07}.
  However, our approach will be to presume, not static knowledge of the type, but dynamic knowledge of $f$'s support.
}

Logic programming implements existential quantification by rejecting input-out\-put directionality:
much as a function is a set of input-output tuples, a predicate can enumerate its tuples---or rather, those whose output is true (any non-enumerated inputs are implicitly false).
We call these inputs the \emph{support} of a predicate.
If we know $f$'s support, calculating $\name{exists}\<f$ is simple: is the support nonempty?

This suggests we can make the direct functional translation of logic programs behave correctly if we equip functions with 
their support.
Inspired by 
Datalog, we focus on {finite} support.
Since finitely supported functions can be represented as key-value tables, we also call them \emph{finite maps.}
\todolater{maybe I should discuss the `goal' here: a language with procedural higher order functions, but also finite maps which are guaranteed finitely supported, represented as tables, and computed like database queries. should this go after or before contribution list? I guess it's a nice pr\'ecis of \fslang.}
We make the following contributions:
\begin{enumerate}
\item We observe that defining a function's support requires a point in its codomain, suggesting the category \Point\ of pointed sets and point preserving maps is a suitable site for the semantics of a functional logic programming language.

\item We show \Point\ forms a model of relevance, a relaxation of linearity in which variables must be used {at least once} rather than {exactly once.}

\item We show finite maps form a graded monad/comonad on \Point.

\item We construct a simply typed language \fslang\ that can express relational algebra and some forms of aggregation, whose type system guarantees finite support.
  \todolater{but we still support nonfinite possibly higher order functions! so we handle the \emph{interaction} between finite \& nonfinite things.}

\item We give a denotational semantics of \fslang\ in \Point. \TODO{cut semantics figure and rephrase as ``\emph{sketch} a denotational semantics of \fslang\ in \Point.''?}
\end{enumerate}




\noindent
Along the way we will observe other curious connections:
\begin{enumerate}
\item Database inner joins are point preserving maps out of the smash product $A \otimes B$ of pointed sets, while outer joins map out of the direct product $A \with B$.
\item Input/output modes in bottom-up logic programming correspond to procedural functions and finite maps respectively.
\item Checking a rule is well moded corresponds to an effect system for the graded monad of finite maps.
\end{enumerate}

\begin{center}
\noindent\begin{CJK*}{UTF8}{min}
さあ始めましょう！ 
\end{CJK*}
\qquad
Let's get started!
\end{center}


\section{Support, pointedly}

\label{sec:support}

The support of a boolean function $f$ is the set $\{ x \in A : f(x) \ne \false \}$.
It will be useful to generalize beyond booleans,
firstly to expose compositional structure, for instance, to define the support of a curried function $f : A \to B \to \bool$;
and secondly to generalize from boolean logic and existential quantification (or relational algebra and projection, respectively) to weighted logic programming and aggregation (or tensor algebra and contraction).

To this end we write $P,Q,R$ for sets with a designated element (`pointed sets') and $\nil_P \in P$ for the designated element (the `point').
(We use $A,B,C$ for ordinary sets.)
For instance, we regard $\bool$ as pointed with $\nil_\bool = \false$.
We define the support of a function $f : A \to P$ into a pointed set to be $\support f = \{ x \in A : f(x) \ne \nil_P \}$.
If one regards $\nil_P$ as an uninteresting, default value, a function's support contains those inputs with interesting, nondefault outputs.

We write $A \fto P = \{ f \in A \to P : \support f ~\text{is finite} \}$ for the finitely supported functions from $A$ to $P$.
This is a pointed set with $\nil_{A \fto P} = \fn{x} \nil_P$;
the boring, default finite map is the one with empty support.
What can we do with finite maps?
We can of course apply them---this is lookup in a key-value table.
We can also curry and uncurry them: $A \times B \fto P \cong A \fto (B \fto P)$.
This converts between a single flat table $A \times B \fto P$ and a trie-like nesting of tables $A \fto B \fto P$.

But how can we manipulate the output $P$ of a finite map $A \fto P$?
Can we, for instance, compose finite maps?
Unfortunately, the composition of two finite maps may not have finite support:
e.g.\ take $\fn{x} x < 3 : \nat \fto \bool$ followed by $\name{not} : \bool \fto \bool$.
The set of naturals less than three is finite; its complement is not.
For the same reason, we cannot compose a finite map $g : A \fto P$ with an arbitrary function $f : P \to Q$ and expect the result to be finitely supported.

However, we \emph{can} compose a finite map with a function $f$ if $f(\nil_P) = \nil_Q$; this preserves finiteness because it can only contract our support.
%
These are the \emph{point preserving maps}, which we notate $P \lto Q = \{ f \in P \to Q : f(\nil_P) = \nil_Q \}$.
These are pointed with $\nil_{P \lto Q} = \fn{x} \nil_Q$.
The category \Point\ has pointed sets $P,Q$ as objects and point preserving maps as morphisms; what we have just observed is that, for any set $A$, there is a functor $F_A : \Point \to \Point$ given by $F_A(P) = A \fto P$.
If finitely supported maps are this work's \emph{raison d'\^etre,} point preserving maps are how we will manipulate and combine them.






\subsection{Direct or/and smash}

Following logic programming, we take particular interest in two point preserving maps: boolean disjunction and conjunction.
Let's start with their types.
Recall that $\nil_\bool = \false$.
Observe that \emph{or} yields \nil\ when both its arguments are \nil, while \emph{and} yields \nil\ when either argument is.
We can reflect this both/either distinction using distinct types.
\noindent
Disjunction accepts
a pair which is \nil\ when both components are.
This is the {direct product} of pointed sets, $P \with Q = \{\withpair{p,q} : p \in P, q \in Q\}$ where $\nil_{P \with Q} = \withpair{\nil_P, \nil_Q}$.
%
Conversely, conjunction accepts
a pair which is \nil\ if either component is.
This is the {smash product} of pointed sets, $P \otimes Q = \{(p,q) : p \in P, q \in Q\}$ quotiented by \(\nil_{P \otimes Q} = (\nil_P, y) = (x, \nil_Q)\).
Thus:

\begin{align*}
  \kword{or} &: \bool \with \bool \lto \bool
  &
  \kword{and} &: \bool \ox \bool \lto \bool
\end{align*}

\noindent
It will be useful to generalize `and' to type $\bool \otimes P \lto P$, along with its mirror image `\whenkword' of type $P \otimes \bool \lto P$, defined:\footnotemark{}
\nopagebreak[2]

\[
\begin{array}{rclcl}
  \true \tand x
  &=& x \when \true
  &=& x
  \\
  \false \tand x
  &=& x \when \false
  &=& \nil_P
\end{array}\]


\footnotetext{
  An edifying exercise is to verify the generalized `and' is one leg of an isomorphism $\bool \otimes P \cong P$; its inverse is $\fn{x} (\true, x) : P \lto \bool \otimes P$.
}

\noindent
Finally, we must note an important asymmetry:
curried maps $f : P \lto Q \lto R$ must preserve \nil\ in each argument separately, like `and' and unlike `or':
\nopagebreak[3]

\begin{align*}
  (f \<\nil_P) \<q &= \nil_{Q \lto R} \<q = \nil_R
  && \text{since $f : P \lto (Q \lto R)$ preserves \nil}\\
  (f \<p) \<\nil_Q
  &=
  \nil_R
  && \text{since $f\<p : Q \lto R$ preserves \nil}
\end{align*}

\noindent
In fact, currying and uncurrying of point preserving maps forms an isomorphism $P \ox Q \lto R \cong P \lto Q \lto R$.
But there is no corresponding way to curry maps $P \with Q \lto R$ (like `or') that yield \nil\ only when both arguments are \nil.\footnote{Phrased categorically, \Point\ does not have all exponential objects.}









\section{A finitely supported map from examples to booleans}

\label{sec:examples}

\begin{figure}[t]
  \begin{center}
    \begin{tabular}{llll}
      \emph{Notation}
      & \emph{Name} & \emph{Elements} & \nil\\\midrule
      $P \with Q$
      & direct product
      & $\withpair{p,q}$ for $p\in P, q \in Q$
      & $\withpair{\nil_P, \nil_Q}$
      \\
      $P \ox Q$
      & smash product
      & \((p,q)\) for \(p \in P, q \in Q\) modulo:
      & \((\nil_P, q) = (p, \nil_Q)\)
      \\
      $P \lto Q$
      & point preserving maps
      & $f : P \to Q$ with $f(\nil_P) = \nil_Q$
      & $\fn{\pwild} \nil_Q$
      \\
      $A \fto P$
      & finite maps
      & $f : A \to P$ with $\support f$ finite
      & $\fn{\pwild} \nil_P$
      \\
      $\bool$
      & booleans
      & \true, \false
      & \false
      \\
      $\natz$
      & natural numbers
      & 0, 1, 2, 3, 4...
      & 0
      \\
      $\maybeof A$
      & maybe type
      & $\tjust a$ for $a \in A$, \cnil
      & \cnil
    \end{tabular}
  \end{center}
  \vspace{-0.25\baselineskip}
  \caption{Pointed sets I have known and loved}
  \label{fig:pointed-sets}
  \vspace{0.5\baselineskip}     
\end{figure}

\begin{figure}[t]
  \[
  \begin{array}{rccll}
    \text{types} & A,B &\bnfeq&
    \typeforget P \bnfor A \to B \bnfor A \times B \bnfor 1 \bnfor ...\\
    \text{pointed types} & P,Q &\bnfeq&
    P \with Q \bnfor P \ox Q \bnfor P \lto Q \bnfor A \fto Q
    \bnfor \maybeof A
    \bnfor \natz
    \\
    \text{expressions} & e &\bnfeq&
    \termforget t
    \bnfor x \bnfor \fn x e \bnfor e_1 \<e_2
    \bnfor (e_1, e_2) \bnfor \pi_i \<e \bnfor ()
    \\&&&
    \ifjust{e_1}{x}{e_2}{e_3}
    \bnfor ...
    \\
    \text{pointed terms} & t,u &\bnfeq&
    \nil
    \bnfor x \bnfor \fn{x} t \bnfor t\<u
    \bnfor t\<x \bnfor t\<e
    \bnfor \withpair{t, u} \bnfor \pi_i \<t
    \\
    &&& (t,u) \bnfor \letpair x y t u
    \bnfor \tjust e \bnfor \letjust x t u
  \end{array}
  \]
  \caption{Syntax of \fslang}
  \label{fig:syntax}
\end{figure}

\newcommand\mult\times

\begin{figure}[t]
  \newcommand\desugar{\longrightarrow}
  \begin{align*}
    \bool &\desugar \maybeof 1 &
    t \tand u &\desugar \letjust \pwild t u
    \\
    \true &\desugar \tjust () &
    \leteq x t u &\desugar \letin{(x,y) = (t, \true)} (y \tand u)
    \\
    \false &\desugar \nil &
    t \when u &\desugar \leteq x t {(u \tand x)}
    \\
    \kword{or} &: \bool \with \bool \lto \bool
    &
    \name{exists} &: (A \fto \bool) \lto \bool
    \\
    (+) &: \natz \with \natz \lto \natz
    &
    \name{sum} &: (A \fto \natz) \lto \natz
    \\
    (\mult) &: \natz \ox \natz \lto \natz &
    ({=}) &: A \to (A \fto \bool)
  \end{align*}
  \caption{Syntax sugar and primitive functions in \fslang}
  \label{fig:primitives}
\end{figure}

Now that we've developed enough notation (summarized in \cref{fig:pointed-sets}, along with a few more pointed sets we will introduce later), we can consider some example programs and their types.
For now we rely on intuition to see that they are semantically well typed; in \cref{sec:relevance,sec:finite-support-context,sec:typing-rules} we will develop typing rules.
First, let's consider some basic uses of conjunction (inner joins, in database parlance), starting with the simplest, cross product:

\begin{code}
  \name{cross} : (A \fto \bool) \lto (B \fto \bool) \lto (A \fto B \fto \bool)\\
  \name{cross} \<f \<g \<x \<y = f\<x \tand g\<y
\end{code}

\noindent
%
%
The type of \name{cross} captures something important:
curried functions preserve \nil\ in all arguments separately, and since $\nil_{A \fto \bool} = \fn{\pwild} \false$ is the 
empty relation, we know from its type alone that the cross product of an empty set with any other relation is empty.
Inner joins in general have this property because they use conjunction.
For instance, intersection:

\begin{code}
  \name{intersect} : (A \fto \bool) \lto (A \fto bool) \lto (A \fto \bool)\\
  \name{intersect} \<f \<g \<x = f\<x \tand g\<x
\end{code}

\noindent
This intersects two finite sets, but we can also more generally `intersect' (i.e.\ filter) a finite set with an arbitrary function $A \to \bool$: \todo{THIS TYPE NO MAKE SENSE, IT REQUIRES $(A \to \bool)$ to be a pointed type, which is semantically fine but we haven't done it yet}

\begin{code}
  \name{filter} : (A \fto \bool) \lto (A \to \bool) \lto (A \fto \bool)\\
  \name{filter} \<f \<g \<x = f\<x \tand g\<x
\end{code}

\noindent
We can also define the intersection of two arbitrary functions $f,g : A \to \bool$, but this will only yield another function, not a finite map.

Let's move on to considering what we can express with the function $\name{exists} : (A \fto \bool) \lto \bool$, now justified by the finite support of its argument:



\begin{code}
  \name{costars} : \name{person} \fto \name{person} \fto \bool\\
  \name{costars} \<x \<y =
  \name{exists} \<(\fn{\name{film}} \name{stars} \<\name{film} \<x
  \tand \name{stars} \<\name{film} \<y)
\end{code}

\noindent
Functions of the shape $(A \fto P) \lto P$ represent \emph{aggregations} into $P$.
For instance, $\name{sum} : (A \fto \natz) \lto \natz$ lets us count the number of films someone has starred in:

\begin{code}
  \name{filmCount} : \name{person} \fto \nat_0\\
  \name{filmCount} \<\name{actor} = \name{sum} \<(\fn{\name{film}}
    1 \when {\name{stars} \<\name{film} \<\name{actor}}
  )
\end{code}

\noindent
In general, for any commutative monoid of the form $(P, {\oplus}, \nil_P)$, since $\nil \oplus \nil = \nil$ we have $\oplus : P \with P \lto P$.
This extends to an aggregation $\bigoplus_P : (A \fto P) \lto P$ that takes a finite map $f : A \fto P$ to the monoid sum $\bigoplus_{x \in \support f} f(x)$.
For instance, the aggregation of the monoid $(\bool, \name{or}, \false)$ is \name{exists}, while the aggregation of $(\natz, +, 0)$ is \name{sum}.

Just as \name{exists} combined with $\name{and} : \bool \ox \bool \lto \bool$ gave us relational composition in \name{costars}, summation combined with $\mult : \natz \ox \natz \lto \natz$ (which holds since $0 \mult y = x \mult 0 = 0 = \nil_{\natz}$) gives us matrix multiplication:

\begin{code}
  \name{matMul} : (A \times B \fto \nat_0) \lto
  (B \times C \fto \nat_0) \lto
  A \times C \fto \nat_0\\
  \name{matMul} \<m \<n \<i \<k = \name{sum} \<(\fn j m\,i\,j \mult n\,j\,k)
\end{code}

\noindent
We have already seen that $A \fto P$ is functorial in $P$.
Given a commutative monoid aggregation over $P$, it is also functorial in the key space $A$, by aggregating the values of keys that collide under some map $f : A \to B$. For \natz\ under \name{sum} this is:

\begin{code}
  \name{map} : (A \to B) \to (A \fto \nat_0) \lto (B \fto \nat_0)\\
  \name{map} \<f \<\name{count} \<b =
    \name{sum}
    \<(\fn{a}
      \name{count}\<a \when \teq{f\<a}{b}
    )
\end{code}

\noindent
Of course, finite sets $A \fto \bool$ and bags $A \fto \natz$ are not only functors but monads.
If besides an aggregation \name{exists}/\name{sum} we have a monoid $(P, {\oast} : P \otimes P \lto P, \boldsymbol 1)$---for \bool\ this is conjunction $(\bool, \name{and}, \true)$, for \natz\ it is multiplication $(\natz, \mult, 1)$---then we can define monadic pure and join:
\todo{presumably this is a lawful monad if this forms a semiring?}

\begin{align*}
  &
  \name{pure} : A \to (A \fto \natz)
  &&
  \name{join} : ((A \fto \natz) \fto \natz) \lto (A \fto \natz)
  \\[-3pt]
  &
  \name{pure} \<x \<a = 1 \when x = a
  &&
  \name{join} \<\name{nested} \<a =
  \name{sum} \<(\fn t \name{nested} \<t \mult t\<a)
\end{align*}




\noindent
We might even dare to consider defining finitely supported predicates recursively:

\begin{code}
  \name{connected} : \name{person} \fto \name{person} \fto \bool\\
  \name{connected} \<x \<y = \name{costars} \<x \<y
  \tor \name{exists}\< (\fn{z} \name{connected} \<x \<z \tand \name{connected} \<z \<y)
\end{code}

\noindent
Unfortunately we will not be able to give typing rules or semantics for this kind of recursion here.
Ensuring the existence of a fixed point would seem to require some sort of monotonicity \`a la Datafun~\cite{datafun} or $\lambda_\vee$~\cite{DBLP:journals/pacmpl/RiouxZ25}, and ensuring it remains finite will be even more difficult; we leave this to future work.

Finally, let's turn our attention to the prototypical outer join: union.

\begin{code}
  \name{union} : (A \fto \bool) \with (A \fto \bool) \lto (A \fto \bool)\\
  \name{union} \<\name{fg} \<x =
    {\pi_1 \<\name{fg} \<x} \tor {\pi_2 \<\name{fg} \<x}
\end{code}

\noindent
We saw in \cref{sec:support} why \name{union}, being a function on $\with$ pairs, cannot be curried.
But why can't we destructure the pair $\name{fg}$?
To explain this, we must understand the variable usage discipline that ensures functions preserve \nil.

\colorlet{invalid}{BrickRed}




\section{The relevance of being relevant} 
\label{sec:relevance}

We've seen that point preserving maps can be applied to the outputs of finitely supported maps while preserving finite support, and allow us to combine multiple maps via inner joins (maps out of $\otimes$ pairs) and outer joins (maps out of $\with$ pairs).
How can we check that maps preserve \nil?
To gain intuition, let's look at a few simple examples:

\begin{align*}
  \name{id} &= \fn{x} x
  \phantom{\fn{x} (x,x)} \negphantom{\fn{x} x}
  : P \lto P
  && \text{\color{DarkGreen}\scshape\ding{51}}
  \\
  \color{invalid}
  \name{three} &
  \color{invalid}
  = \fn{x} 3
  \phantom{\fn{x} (x,x)} \negphantom{\fn{x} 3}
  : P \lto \natz
  && \text{\color{invalid}\scshape\ding{55}\: not point preserving}
  \\
  \name{dup}_{\ox} &= \fn{x} (x,x)
  : P \lto P \ox P
  && \text{\color{DarkGreen}\scshape\ding{51}}
  \\
  \name{dup}_{\with} &= \fn{x} \withpair{x,x}
  : P \lto P \with P
  && \text{\color{DarkGreen}\scshape\ding{51}}
\end{align*}

\noindent
Plainly the identity function preserves \nil, but constant functions do not (except for $\fn x \nil$).
This suggests a linear type system, which ensures each variable is used exactly once;
constant functions do not use their argument and so are prohibited.
However, duplication does preserve \nil, whether into $\ox$ or $\with$ pairs: $(\nil, \nil) = \nil$ and $\withpair{nil,nil} = \nil$.
Or, consider the intersection $(\fn x f\<x \tand g\<x)$ of two maps $f,g : P \lto \bool$, which uses $x$ twice yet still preserves \nil.
What we need is a \emph{relevant} type system, which ensures variables are used \emph{at least} once---although as we'll see presently, what counts as `used' can be subtle.

So far our examples fail to distinguish the behavior of $\ox$ from $\with$; let's fix that:

\begin{align*}
  \color{invalid}
  \name{fst}_{\ox} &
  \color{invalid}
  = \fn{p} \letpair x y p x
  \color{invalid}
  : P \ox Q \lto P
  && \text{\color{invalid}\scshape\ding{55}\: not point preserving}
  \\
  \name{fst}_{\with} &= \fn{p} \pi_1 \<p
  \phantom{\fn{p} \letpair x y p x}
  \negphantom{\fn p \pi_1 \<p}
  : P \with Q \lto P
  && \text{\color{DarkGreen}\scshape\ding{51}}
  \\
  \name{pair}_{\ox} &= \fn{x} \fn{y} (x,y)
  \phantom{\fn{p} \letpair x y p x}
  \negphantom{\fn x \fn y (x,y)}
  : P \lto Q \lto P \ox Q
  && \text{\color{DarkGreen}\scshape\ding{51}}
  \\
  \color{invalid}
  \name{pair}_{\with} &
  \color{invalid}
  = \fn{x} \fn{y} \withpair{x,y}
  \phantom{\fn{p} \letpair x y p x}
  \negphantom{\fn x \fn y \withpair{x,y}}
  \color{invalid}
  : P \lto Q \lto P \with Q
  && \text{\color{invalid}\scshape\ding{55}\: not point preserving}
  \\
  \name{and3}_{\ox} &= \fn{x} (x, 3)
  \phantom{\fn{p} \letpair x y p x}
  \negphantom{\fn{x} (x,3)}
  : P \lto P \ox \natz
  && \text{\color{DarkGreen}\scshape\ding{51}}
  \\
  \color{invalid}
  \name{and3}_{\with} &
  \color{invalid}
  = \fn{x} \withpair{x,3}
  \phantom{\fn{p} \letpair x y p x}
  \negphantom{\fn{x} \withpair{x,3}}
  : P \lto P \with \natz
  && \text{\color{invalid}\scshape\ding{55}\: not point preserving}
\end{align*}

\noindent
One can verify these examples mechanically using the definitions of $\ox$ and $\with$, but some intuition may be helpful.
To use an $\ox$ pair $(x,y)$ we must use both $x$ and $y$, to guarantee that if either is $\nil$ we will propagate it.
By contrast, if a $\with$ pair is $\nil$ then both of its components are, so we are free to use only one.
This is why $\name{fst}_{\ox}$ is invalid but $\name{fst}_{\with}$ is correct---and also why we use destructuring to eliminate $\ox$, but projection for $\with$.
Conversely, constructing an $\ox$ pair $(t,u)$ uses anything used by either $t$ or $u$, since if either $t$ or $u$ is $\nil$ the whole pair is,
but constructing a $\with$ pair $\withpair{t,u}$ uses only what both $t$ and $u$ use, since only these will force both $t$ and $u$ to be \nil.
This is why $\name{pair}_{\ox}$ and $\name{and3}_{\ox}$ are correct but $\name{pair}_{\with}$ and $\name{and3}_{\with}$ are invalid.


Since we need to mix a cartesian, structural type system for sets $A,B$ and functions $A \to B$ with a substructural, relevant type system for pointed sets $P,Q$ and point preserving maps $P \lto Q$, we adapt the rules of Benton and Wadler's mixed linear/nonlinear logic LNL~\cite{DBLP:conf/lics/BentonW96,DBLP:conf/csl/Benton94} to relevance rather than linearity.
We use two contexts, $\G$ containing ordinary variables $\hyp x A$ and $\D$ containing pointed set variables $\hyp x P$; two syntactic classes of terms, ordinary $e$ and point preserving $t$; and two typing judgments, $\G \entails e : A$ for functions and $\G \cxsep \D \entails t : P$ for point preserving maps.
A few example rules:
\begin{mathpar}
  \infer{
    \lJ t P \G \D\\
    \lJ u Q \G \D
  }{
    \lJ {\withpair{t,u}} {P \with Q} \G \D
  }

  \infer{
    \lJ t P \G {\D_1}\\
    \lJ u Q \G {\D_2}
  }{
    \lJ {(t,u)} {P \ox Q} \G {\D_1 \cup \D_2}
  }

  \infer{~}{\lJ x P \G {\hyp x P}}

  \infer{\lJ t Q {\G} {\D, \hyp x P}}{
    \lJ {\fn x t} {P \lto Q} \G \D
  }
\end{mathpar}

\noindent
Observe that in the rule for $\with$ pairs $\withpair{t,u}$ each component is checked in the same relevant context $\D$, and therefore must use the same variables; but in $\ox$ pairs $(t,u)$, we give them different contexts $\D_1, \D_2$ which must \emph{union} to produce the context of the whole pair.
This union is a key difference from standard linear logic, which would split the contexts disjointly; it allows a variable to be used in both branches---but, unlike the $\with$ rule, does not \emph{require} it.




\section{The other side of the tracks}
\label{sec:finite-support-context}

Alas, these are not yet the typing rules we are looking for.
A scant two contexts will not suffice; we have \emph{three} kinds of function we'd like to introduce---ordinary $A \to B$, point preserving $P \lto Q$, and finitely supported $A \fto P$---and therefore three kinds of variable, needing three separate contexts.
Besides $\G$ and $\D$ we need a context $\O$ of finitely supported variables $\hyp x A, \hyp y B, ...$, and our typing judgement takes the form $\J t P \G \D \O$.
To understand how $\O$ behaves,
let's revisit the examples that opened the previous section:

\begin{align*}
  \color{invalid}
  \name{id} &
  \color{invalid}
  = \fn{x} x
  \phantom{\fn{x} (x,x)} \negphantom{\fn{x} x}
  : P \fto P
  && \text{\color{invalid}\scshape\ding{55}\: not finitely supported}
  \\
  \color{invalid}
  \name{three} &
  \color{invalid}
  = \fn{x} 3
  \phantom{\fn{x} (x,x)} \negphantom{\fn{x} 3}
  \color{invalid}: A \fto \natz
  && \text{\color{invalid}\scshape\ding{55}\: not finitely supported}
  \\
  \color{invalid}
  \name{dup}_{\ox} &
  \color{invalid}
  = \fn{x} (x,x)
  \color{invalid}
  : P \fto P \ox P
  && \text{\color{invalid}\scshape\ding{55}\: not finitely supported}
  \\
  \color{invalid}
  \name{dup}_{\with} &
  \color{invalid}
  = \fn{x} \withpair{x,x}
  : P \fto P \with P
  && \text{\color{invalid}\scshape\ding{55}\: not finitely supported}
\end{align*}

\noindent
Oh no!
What went wrong?
Well, constant functions like \name{three} are not finite because their support is their entire possibly infinite domain ($\fn{x} \nil$ is the only exception).
Similarly, the support of \name{id}, $\name{dup}_{\ox}$, and $\name{dup}_{\with}$ is the entire domain minus \nil; we cannot generally use a finite map's input directly in its output.\footnotemark{}

\footnotetext{We could in principle make a carve-out 
  for maps with finite domains.
  As we intend to represent finite maps by tables, however, this poses a usability hazard: enumerating all 64-bit integers just because a programmer wrote $\fn{x} x$ is not desirable.
}

If $\fn{x} x$ is not well typed, what does our variable rule look like?
How do we use variables in our finitely supported context $\O$?
Let's return to our source of inspiration.
In bottom-up logic programming, we generate nonempty, finite relations by either (1) combining other nonempty relations or (2) using rules that refer to constants.
We saw in \cref{sec:examples} that in \fslang, (1) means using point preserving maps to transform finite maps, and (2) means using equality $(=) : A \to A \fto \bool$ to generate singleton finite maps.
Let's consider a prototypical example of each kind:

\begin{align*}
  \name{actorOrDirector} &= \fn{x} \name{actor} \<x \tor \name{director} \<x
  : \name{person} \fto \bool
  && \text{\color{DarkGreen}\scshape\ding{51}}
  \\
  \name{hitchcockAlone} &= \fn{x} \name{hitchcock} = x
  \phantom{\fn{x} \name{actor} \<x \tor \name{director} \<x}
  \negphantom{\fn{x} \name{hitchcock} = x}
  : \name{person} \fto \bool
  && \text{\color{DarkGreen}\scshape\ding{51}}
\end{align*}

\noindent
In each case, we use the finitely supported variable $x$ by applying a finite map to it.
This, then, will serve both as our elimination rule for finite maps and our usage rule for finitely supported variables:
\begin{mathpar}
  \infer*[right=$\fto$\,i]{
    \J t P \G \D {\O, \hyp x A}
  }{
    \J {\fn x t} {A \fto P} \G \D \O
  }
  
  \infer*[right=$\fto$\,e]{
    \J t {A \fto P} \G \D \O
  }{
    \J {t\<x} P \G \D {\O, \hyp x A}
  }
\end{mathpar}

\noindent
These rules structuralize the isomorphism $A \fto B \fto P \cong A \times B \fto P$ (which is more obvious after renaming: $\O \fto A \fto P \cong \O \times A \fto P$).
Along with the isomorphism $P \cong (1 \fto P)$, this shows that finitely supported maps form not just a functor but a \emph{graded monad} on \Point.

\todolater{I'd really like to do this connection more justice.}
A (non-graded) monad is an endofunctor $F$ with natural transformations $\name{pure}_\alpha : \alpha \to F\alpha$ and $\name{join}_\alpha : FF\alpha \to F\alpha$, satisfying laws which we omit for brevity.
For our purposes, a graded monad~\cite{Smirnov08,10.1145/2535838.2535846} is a family of functors $F_m$ where $m \in M$ is drawn from some monoid of grades $(M, {\cdot}, 1)$, with analogues of \name{pure} and \name{join} that interact with the grading monoid: $\name{pure}_\alpha : \alpha \to F_1 \alpha$ and $\name{join}_\alpha : F_m F_n \alpha \to F_{mn}\alpha$, again satisfying certain omitted laws.

In our case, the graded monad is $F_A P = A \fto P$, and the monoid of grades is sets under cross product.\footnotemark{}
This makes $\name{pure}_P : P \lto (1 \fto P)$ and $\name{join}_P : (A \fto B \fto P) \lto (A \times B \fto P)$ merely the forward legs of our two isomorphisms.
The laws we omitted hold trivially because these maps are isomorphisms---indeed, the isomorphisms' reverse legs also make $F_A$ a graded comonad~\cite{DBLP:conf/icfp/PetricekOM14}.
Although almost trivial, this graded (co)monad is of practical interest because type systems for graded monads (effect systems) have been extensively studied~\cite{10.1145/3022670.2951939}.
\todolater{mention ours is a bit weird in that the effect is a context. which makes you think it might be a coeffect system, but it's not, because ... WHY?}

\footnotetext{Technically this is not a monoid, since $(A \times B) \times C$ and $A \times (B \times C)$ are only isomorphic, not equal.
  This technicality is not worth getting hung up on; it can be defeated, for instance, by taking grades to be contexts $\O$ under concatenation rather than sets.}


Unfortunately, our typing rules do not all follow neatly from this connection to grading.
Let's examine the rules for $\with$ and $\otimes$.
Both are informative: the former because it is straightforward, the latter because it is not. The rules for $\with$ are:
\begin{mathpar}
  \infer*[right=$\with$\,i]{
    \J t P \G \D \O
    \\
    \J u Q \G \D \O
  }{
    \J {\withpair{t,u}} {P \with Q} \G \D \O
  }

  \infer*[right=$\with$\,e]{
    \J t {P_1 \with P_2} \G \D \O
  }{
    \J {\pi_i\<t} {P_i} \G \D \O
  }
\end{mathpar}

\noindent
%
Now let's
consider what rules we would need to type the following examples involving $\otimes$ (by way of $\name{and} : \bool \ox \bool \lto \bool$):
\begin{align*}
  {\hyp f {A \fto \bool}, \hyp g {A \fto \bool}} \cxsep
  \emptycx \cxsep
  {\hyp x A, \hyp y A}
  &\mathbin{\,\vdash} {f\<x \tand g\<y} : {\bool}\\
  {\hyp f {A \fto \bool}, \hyp g {A \fto \bool}} \cxsep
  \emptycx \cxsep
  {\hyp x A \phantom{, \hyp y A}}
  &\mathbin{\,\vdash} {f\<x \tand g\<x} : \bool\\
  {\hyp f {A \fto \bool}, \hyp g {A \to \bool}} \cxsep
  \emptycx \cxsep
  {\hyp x A \phantom{, \hyp y A}}
  &\mathbin{\,\vdash} {f\<x \tand g\<x} : \bool
\end{align*}


\noindent
Since we wish to guarantee finite support, we first ask:
what is the support of each example (as a function of the supports of $f, g$)?

The support of $f\<x \tand g\<y$ is the cross product of the supports of $f : A \fto \bool$ and $g : A \fto \bool$.
Fortunately, the cross product of finite sets is finite.
So what typing rule does this example need?
We have used each variable $x,y$ in our finite support context $\O$ exactly once, so it suffices to be able to split this context between the (implicit) $\ox$ pair of arguments to `and':

\[
  \infer*{
    \J t P \G {\D_1} {\O_1}
    \\
    \J u Q {\G} {\D_2} {\O_2}
  }{
    \J{(t, u)}{P \ox Q}{\G}{\D_1 \cup \D_2}{\O_1, \O_2}
  }
\]

\noindent
This rule does not suffice for our second example, $f\<x \tand g\<x$, which uses the same finitely supported variable $x$ twice.
Nonetheless this has finite support, namely, the intersection of $f$'s and $g$'s supports.
So we can relax our rules to union contexts, much like the rules for the relevant context $\D$:

\[
  \infer*{
    \J t P \G {\D_1} {\O_1}
    \\
    \J u Q {\G} {\D_2} {\O_2}
  }{
    \J{(t, u)}{P \ox Q}{\G}{\D_1 \cup \D_2}{\O_1 \cup \O_2}
  }
\]

\noindent
Yet this still fails to check our final example, which filters the finite set $f: A \fto \bool$ by an arbitrary boolean function $g: A \to \bool$.
This is unproblematic semantically: we can filter a finite set by any predicate we like and the result remains finite!
The solution is a rule that breaks the rules:
any variable finitely supported by $t$ may be used unrestricted in $u$, so the finite support context $\O_1$ for $t$ will \emph{jump}
across the railway tracks
into the unrestricted context for $u$:

\[
  \infer*[right=$\ox$\,i]{
    \J t P \G {\D_1} {\O_1}
    \\
    \J u Q {\G, \O_1} {\D_2} {\O_2}
  }{
    \J{(t, u)}{P \ox Q}{\G}{\D_1 \cup \D_2}{\O_1, \O_2}
  }
\]

\noindent
This rule is asymmetric: it `grounds' variables from left to right.
This choice is arbitrary---right to left is just as sound---but the symmetric variant, which allows variables finitely supported by either $t$ or $u$ to be used unrestricted in the other, allows circular dataflow and is unsound.
For instance, there may be infinitely many $x,y$ such that $x = y$, but for fixed $x$ there is exactly one such $y$; this is why we give equality the type $A \to (A \fto \bool)$.
However, if we allow each branch to ground variables used in the other, then $(x = y \tand y = x)$, though semantically identical to $x = y$, would incorrectly appear to finitely support both $x$ and $y$: in the left branch, $x$ is used to ground $y$, and vice-versa.

\todolater{discuss how the three different typing rules corresponds to three different implementation strategies: cross products, inner joins, and subqueries!}

To ensure our last rule generalizes the two prior ones, we must be able to weaken a finitely supported hypothesis $\hyp x A \in \O$ to an unrestricted one $\hyp x A \in \G$.
Then a variable finitely supported by both $t$ and $u$ can simply be placed into $\O_1$ in \rulename{$\ox$\,i}.
Unfortunately this doesn't yet hold: \rulename{$\fto$\,e} lets us apply finite maps $A \fto P$ to variables in $\O$ but not in $\G$!
We need an additional rule that lets us apply a finite map to an unrestricted variable $x$, or more generally an expression $e$ which (as usual) may use unrestrictedly any variables finitely supported by $t$:\footnotemark{}

\[
\infer*[right=${\fto}\,\textsc{e}_2$]{
  \J t {A \fto P} \G \D \O\\
  \cJ e {A} {\G,\O}
}{
  \J {t\<e} P \G \D \O
}
\]

\footnotetext{Adding $\O$ to $e$'s context is also needed to allow multiple uses of a finitely supported variable among the arguments to a nested finite map, e.g.\ $\J{f\<x\<x}{P}{\emptycx}{\hyp f {A \fto A \fto P}}{\hyp x A}$.}



\begin{figure}
  \begin{mathpar}
    \boxed{\cJ e A \G}\hfill\\  

    \infer*[right=ui]{\J t P \G \emptycx \emptycx}{\cJ t P \G}

    \infer*[right=evar]{\hyp x A \in \G}{\cJ x A \G}

    \infer*[right=$1$\,i]{~}{\cJ {()} {1} \G}

    \infer*[right=$\to$\,i]{\cJ e B {\G,\hyp x A}}{\cJ {\fn x e} {A \to B} \G}

    \infer*[right=$\to$\,e]{
      \cJ {e_1} {A \to B} \G\\
      \cJ {e_2} {A} \G
    }{\cJ {e_1 \<e_2} B \G}

    \infer*[right=$\times$\,i]{
      \cJ {e_1} A \G\\
      \cJ {e_2} B \G
    }{
      \cJ {(e_1, e_2)} {A \times B} \G
    }

    \infer*[right=$\times$\,e$_i$]{
      \cJ {e} {A_1 \times A_2} \G
    }{
      \cJ {\pi_i \<e} {A_i} \G
    }

    \infer*[right=case]{
      \cJ {e_1} {\maybeof A} \G\\
      \cJ {e_2} B {\G, \hyp x A}\\
      \cJ {e_3} B \G
    }{
      \cJ {\ifjust {e_1} x {e_2} {e_3}} B \G
    }
    \vspace{1em}
    \\
    \boxed{\J t P \G \D \O}\hfill\\ 

    \infer*[right=ue]{\cJ e P \G}{\J e P \G \emptycx \emptycx}

    \infer*[right=var]{~}{\J x P \G {\hyp x P} {\emptycx}}

    \infer*[right=nil]{~}{\J \nil P \G \D \O}
    \\
    \infer*[right=$\lto$\,i]{
      \J t Q \G {\D, \hyp x P} {\emptycx}
    }{
      \J {\fn x t} {P \lto Q} \G \D {\emptycx}
    }

    \infer*[right=$\fto$\,i]{
      \J t P \G \D {\O, \hyp x A}
    }{
      \J {\fn x t} {A \fto P} \G \D \O
    }

    \infer*[right=$\lto$\,e]{
      \J t {P \lto Q} \G {\D_1} {\O_1}
      \\\\
      \J u {P} {\G,\O_1} {\D_2} {\O_2}
    }{
      \J{t\<u}{Q}{\G}{\D_1 \cup \D_2}{\O_1, \O_2}
    }

    \infer*[right=$\fto$\,e]{
      \J t {A \fto P} \G {\D} {\O}
    }{
      \J {t\<x} P \G {\D} {\O, x : A}
    }

    \infer*[right=$\with$\,i]{
      \J t P \G \D \O
      \\\\
      \J u Q \G \D \O
    }{
      \J {\withpair{t,u}} {P \with Q} \G \D \O
    }

    \infer*[right=$\ox$\,i]{
      \J t P \G {\D_1} {\O_1}
      \\\\
      \J u Q {\G, \O_1} {\D_2} {\O_2}
    }{
      \J{(t, u)}{P \ox Q}{\G}{\D_1 \cup \D_2}{\O_1, \O_2}
    }

    \infer*[right=$\with$\,e]{
      \J t {P_1 \with P_2} \G \D \O
    }{
      \J {\pi_i\<t} {P_i} \G \D \O
    }

    \infer*[right=$\ox$\,e]{
      \J t {P \ox Q} \G {\D_1} {\O_1}
      \\\\
      \J u Q {\G, \O_1} {\D_2, \hyp{x}{P}, \hyp{y}{Q}} {\O_2}
    }{
      \J{\letpair{x}{y} t u}{Q}{\G}{\D_1 \cup \D_2}{\O_1, \O_2}
    }


    \infer*[right=${\fto}\,\textsc{e}_2$]{
      \J t {A \fto P} \G \D \O\\
      \cJ e {A} {\G,\O}
    }{
      \J {t\<e} P \G \D \O
    }

    \infer*[right=maybe i]{     
      \cJ e A \G
    }{
      \J {\tjust e} {\maybeof A} \G \emptycx \emptycx
    }

    \infer*[right=maybe e]{     
      \J t {\maybeof A} \G {\D_1} {\O_1}
      \\
      \J u P {\G, \O_1, x : A} {\D_2} {\O_2}
    }{
      \J {\letjust x t u} P \G {\D_1 \cup \D_2} {\O_1, \O_2}
    }
  \end{mathpar}
  \caption{\fslang\ typing rules}
  \label{fig:typing-rules}
\end{figure}

\begin{figure}[p]
  \begin{align*}
    \den{1} &= 1 & \den{A \times B} &= \den{A} \times \den{B} & \den{A \to B} &= \den{A} \to \den{B}
    &
    \den{P} &= U \den{P}
  \end{align*}

  \begin{align*}
    \den{P \with Q} &= \den{P} \with \den{Q} &
    \den{P \ox Q} &= \den{P} \ox \den{Q} &
    \den{P \lto Q} &= \den{P} \lto \den{Q}\\
    \den{A \fto P} &= \den{A} \fto \den{P} &
    \den{\maybeof A} &= \maybeof{\den{A}} &
    \den{\natz} &= \natz
  \end{align*}

  \begin{align*}
    \den{\G} &= \prod_{\hyp x A \in \G} \den{A}
    &
    \den{\D} &= \bigotimes_{\hyp x P \in \D} \den{P}
    &
    \den{\O} &= \prod_{\hyp x A \in \G} \den{A}
  \end{align*}

  \caption{Semantics of types and contexts in \fslang}
  \label{fig:type-context-semantics}
\end{figure}

\newcommand\g\gamma
\renewcommand\d\delta
\renewcommand\o\omega
\newcommand\Den[1]{\left\llbracket{#1}\right\rrbracket}

\begin{figure*}[p]
  \textsc{semantics of $\cJ e A \G$}
  \begin{align*}
    \den{t} \<\g &= \den{t} \<\g \<() \<()
    \\
    \den{x} \<\g &= \pi_x \g
    \\
    \den{()} \<\g &= ()
    \\
    \den{\fn x e} \<\g &= \fn x \den{e} \<(\g,x)
    \\
    \den{e_1 \<e_2} \<\g &= \den{e_1} \<\g \<(\den{e_2} \<\g)
    \\
    \den{(e_1, e_2)} \<\g &= (\den{e_1} \<\g,\, \den{e_2} \<\g)
    \\
    \den{\pi_i\<e} \<\g &= \pi_i (\den e \<\g)
    \\
    \den{\ifjust{e_1} x {e_2} {e_3}} &=
    \begin{cases}
      \den{e_2} \<(\g,x) & \text{if}~ \den{e_1} \<\g = \cjust~x\\
      \den{e_3} \<\g & \text{otherwise}
    \end{cases}
    \intertext{\textsc{semantics of $\J t P \G \D \O$}}
    \den{{\J e P \G \emptycx \emptycx}}
    \<\g \<\d &=
    \{() \mapsto \den{e} \<\g : \d \ne \nil \}
    \\
    \den{\J \nil P \G \D \O} \<\g \<\d &= \{\} 
    \\
    \den{\J x P \G {\hyp x P} \emptycx} \<\g \<x &= \{ () \mapsto x \}
    \\
    \den{\J {\fn x t} {P \lto Q} \G \D {\emptycx}}
    \<\g \<\d &=
    \{ () \mapsto \fn{x} \den{t} \<\g \<(\d,x) \}
    \\
    \den{\J {\fn x t} {A \fto P} \G \D \O}
    \<\g \<\d &=
    \{ \o \mapsto \{ x \mapsto y : (\o, x) \mapsto y \in \den{t} \<\g \<\d \}
    : \ex{x,y} (\o,x) \mapsto y \in \den{t} \<\g \<\d \}
    \\
    \den{\J{t\<u}{Q}{\G}{\D_1 \cup \D_2}{\O_1, \O_2}}
    \<\g \<\d
    &= \{ (\o_1, \o_2) \mapsto x\<y :
    \o_1 \mapsto x \in \den t \<\g \<(\pi_{\D_1} \d),\,
    \o_2 \mapsto y \in \den u \<(\g, \o_1) \<(\pi_{\D_2} \d)
    \}
    \\
    \den{\J {t\<x} P \G {\D} {\O, x : A}}
    \<\g \<\d &=
    \{ (\o, x) \mapsto y :
    \o \mapsto f \in \den{t} \<\g \<\d,\,
    x \mapsto y \in f
    \}
    \\
    \den{\J {t\<e} P \G {\D} {\O}}
    \<\g \<\d &=
    \{ \omega \mapsto f(\den{e}\<(\g,\o)) : \omega \mapsto f \in \den{t}\<\g\<\d \}
    \\
    \den{\J {\withpair{t,u}} {P \with Q} \G \D \O}
    \<\g \<\d &=
    \{ \o \mapsto \withpair{x,y} :
    \o \mapsto x \in \den{t} \<\g \<\d,\,
    \o \mapsto y \in \den{u} \<\g \<\d
    \}
    \\
    \den{\J{(t, u)}{P \ox Q}{\G}{\D_1 \cup \D_2}{\O_1, \O_2}}
    \<\g \<\d &=
    \{ (\o_1, \o_2) \mapsto (x,y) :
    \o_1 \mapsto x \in \den{t} \<\g \<(\pi_{\D_1} \d),\,
    \o_2 \mapsto y \in \den{u} \<(\g,\o_1) \<(\pi_{\D_2} \d)
    \}
    \\
    \den{\J {\pi_i\<t} {P_i} \G \D \O}
    \<\g \<\d &=
    \{ \o \mapsto \pi_i\, x : \o \mapsto x \in \den{t} \<\g \<\d \}
    \\
    \den{\J{\letpair{x}{y} t u}{Q}{\G}{\D_1 \cup \D_2}{\O_1, \O_2}}
    \<\g \<\d &=
    \{ (\o_1,\o_2) \mapsto z :
    \o_1 \mapsto (x,y) \in \den{t} \<\g \<(\pi_{\D_1} \d),\,
    \o_2 \mapsto z \in \den{u} \<(\g, \o_1) \<(\pi_{\D_2} \d, x, y)
    \}
    \\
    \den{\J {\tjust e} {\maybeof A} \G \emptycx \emptycx}
    \<\g \<\d &=
    \{ () \mapsto \cjust \<(\den{e} \<\g) : \d \ne \nil \}
    \\
    \den{\J {\letjust x t u} P \G {\D_1 \cup \D_2} {\O_1, \O_2}}
    \<\g \<\d &=
    \{ (\o_1, \o_2) \mapsto y :
    \o_1 \mapsto \cjust~x \in \den{t} \<\g \<(\pi_{\D_1} \d),\,
    \o_2 \mapsto y \in \den{u} \<(\g,\o_1,x) \<(\pi_{\D_2} \d)
    \}
  \end{align*}
  \caption{Semantics of expressions and terms in \fslang}
  \label{fig:semantics}
\end{figure*}


\section{Types and semantics}
\label{sec:typing-rules}

We give the typing rules of \fslang\ in \cref{fig:typing-rules}.
The attentive reader will find few surprises.
As mentioned in \cref{sec:relevance}, our rules borrow from adjoint calculi such as LNL~\cite{DBLP:conf/lics/BentonW96,DBLP:conf/csl/Benton94} to handle the interaction between ordinary functions and point preserving functions.
This corresponds semantically to an adjunction between the cartesian closed category $\Set$ (LNL's $\mathcal{C}$) and the symmetric monoidal closed category $\Point$ (LNL's $\mathcal{L}$).
Additionally, \Point\ possesses a natural family of diagonal maps $\fn{x} (x,x) : P \lto P \ox P$, making it a relevant monoidal category~\cite{DBLP:journals/sLogica/Petric02,relevant-categories-and-partial-functions}.
The adjunction in question consists of the free functor $\maybe : \Set \to \Point$ and the forgetful functor $U : \Point \to \Set$.
\todolater{more details/intuition?}
Our main difference from a standard adjoint calculus is our additional context $\O$ of finitely supported variables, which in most typing rules follows the context-hopping pattern of $\ox$ introduction that we explored in \cref{sec:finite-support-context}.

Unsurprisingly, we interpret `set' types as sets $\den{A} \in \Set$, and `pointed set' types as pointed sets $\den{P} \in \Point$.
We have deliberately used almost identical notation for the syntax of types and for the mathematical objects they denote.
See \cref{fig:type-context-semantics} for the semantics of types and contexts.
The $n$-ary smash product $\bigotimes_i P_i$ deserves some explanation: this denotes the pointed set $\{ (x_1, \dots, x_n) : x_1 \in P_1, \dots, x_n \in P_n \} \uplus \{\nil\}$ quotiented by $\fa{i} x_i = \nil \implies (x_1, ..., x_n) = \nil$.
The explicit addition of $\nil$ means that the nullary smash product, $\bigotimes \{\}$, has two elements: $\nil$ and the empty tuple $()$, which is important for interpreting the case $\D = \emptycx$ of an empty pointed set context. \todolater{more intuition}

\begin{align*}
  \den{\cJ e A \G} &\in \den{\G} \to \den{A}\\
  \den{\J t P \G \D \O} &\in \den{\G} \to \den{\D} \lto \den{\O} \fto \den{P}
\end{align*}

\noindent
We give semantics to expressions $e$ and terms $t$ in \cref{fig:semantics}.
The term semantics often depend on the splitting of the $\D$ and $\O$ contexts, so we include the conclusion of the typing judgment in the semantic brackets.
For space reasons we could not include the premises, so cross-referencing with \cref{fig:typing-rules} is suggested.
To make it clear that terms are finitely supported with respect to the context $\O$, we construct all finite maps explicitly using set comprehensions as sets of input-output pairs, $x \mapsto y$. \todolater{this section is underbaked}








\section{Looking back and forward}

There are many ways to combine logic and functional programming.
Mercury~\cite{DBLP:journals/jlp/SomogyiHC96} integrates
functional programming into top-down logic programming.
Functional IncA~\cite{DBLP:conf/ecoop/PacakE22} embeds functional programming into Datalog via a demand transform.
Flix~\cite{10.1145/3428193} has two sublanguages, logical and functional, which can each invoke the other.

\fslang, however, belongs to the subfamily of languages that integrate logical features into a functional substrate.
Perhaps its best known exemplars are `functional logic programming' languages like Curry~\cite{curry} and Verse~\cite{DBLP:journals/pacmpl/AugustssonBCJJSSS23}, 
where relations $A \times B \to \bool$ become `functions' $A \to B$ that may yield many $B$s for a given $A$.
We find this undesirable because it (a) makes multiple returns and unification into pervasive ambient effects and (b) hides the boolean-ness, losing the opportunity to generalize from \bool\ to other types.
\fslang\ is closer to Datafun~\cite{datafun} and $\lambda_\vee$~\cite{DBLP:journals/pacmpl/RiouxZ25}, which avoid ambient effects by representing relations as finite sets, a separate type from normal functions.
However, they express logical operations via loops or set comprehensions.
In contrast \fslang\ is inspired by 
Rel~\cite{10.1145/3722212.3724450},
where relations are defined pointwise like functions,
using `and', `or', and `exists'.
Rel also supports arithmetic operators and aggregations, as if relations had return values.
However, Rel implements this by desugaring to a Datalog-like core where the `value' of a Rel expression is just the relation's final column.
\fslang\ reimagines this syntactic sugar semantically.

By generalizing beyond booleans, \fslang\ moves closer to weighted logic programming languages like Dyna~\cite{filardo:dyna2,matthewfl:thesis}, ProbLog~\cite{DBLP:conf/ijcai/RaedtKT07}, or Datalog\textdegree~\cite{DBLP:journals/jacm/KhamisNPSW24}.
Indeed, \fslang\ is inspired by and indebted to databases research on $K$-relations~\cite{DBLP:conf/pods/GreenKT07}, generalizing Datalog semantics from booleans to semirings, which has in turn inspired functional query languages~\cite{DBLP:journals/corr/abs-2207-00850,DBLP:journals/pacmpl/ShaikhhaHSO22}.
Other Datalog-inspired languages like Datafun, Flix, and $\lambda_\vee$ generalize to semilattices.
However, so far as we know only \fslang\ uses {pointed sets}---the minimal, most general structure required to define a function's support.
Semirings then emerge as a natural structure over pointed sets:
we can generalize \name{or} and \name{exists} to an `additive' commutative monoid $(P, {\oplus} : P \with P \lto P, \nil_P)$ and its aggregation operator $\bigoplus : (A \fto P) \lto P$, and similarly \name{and} generalizes to a `multiplicative' monoid $(P, {\oast} : P \otimes P \lto P, \boldsymbol 1)$ distributing over $\oplus$.

\TODO{discuss connection between my work and Morel's ``predicate inversion''?}

\todolater{Discuss input/output modes.
  Cite Mercury, Mistral Contrastin~\cite{DBLP:conf/ppdp/ContrastinOR18}, and Ariadne Si Suo's paper with Neel K (unpublished!).
  This idea is latent in languages like Datafun, Flix, and $\lambda_\vee$ that separate procedural functions from logical predicates.
  But, although we have barely discussed it in this paper, our observation that grounding corresponds to a effect graded by the set of grounded variables seems novel (so far as we know).
}



\fslang\ has many limitations which we would like to lift in future work:

\begin{description}
\item[Recursion]
  There are two kinds of recursion we might want in \fslang:
  lazy functional recursion (as in recursively defined functions or codata) and bottom-up iteration until reaching a fixed point (as in Datalog's recursively defined relations).
  It's not obvious how these should interact semantically. $\lambda_\vee$~\cite{DBLP:journals/pacmpl/RiouxZ25} shows how to unify them, but requires a lot of structure (semilattices, dcpos, and continuity); moreover, it seemingly cannot do stratified computation, where we compute a fixed point and then use its results in a non-monotone way.


\item[Implementation]
  We did not show how to implement either our type system or our semantics, although
  we have built \href{https://github.com/rntz/fslang/blob/b2730369f68f1cdb4fd69b4a8d997b167d29b7a0/fslang.rkt}{a prototype} (\url{https://github.com/rntz/fslang/blob/main/fslang.rkt}) which we lack space to discuss here.
  As our finite map operations are a form of tensor algebra, perhaps indexed streams~\cite{10.1145/3591268} could provide an efficient implementation technique.

\item[Interleaving $\boldsymbol\D/\boldsymbol\O$ contexts]
  We use three contexts, $\G \cxsep \D \cxsep \O \entails P$, glossed as $\G \to \D \lto \O \fto P$.
  This is not enough: since $P \lto (A \fto Q) \ncong A \fto (P \lto Q)$, writing programs of the latter type naturally requires another context, $\G \cxsep \D_1 \cxsep \O \cxsep \D_2 \entails P$.
  In general we may need any number of interleavings of $\D$ and $\O$ contexts.
  Since our contexts 
  handle the interaction of a graded effect ($\O$ for finite maps) and a coeffect ($\G$ for the \maybe\ comonad) on the category \Point\ ($\D$),
  we hope to apply or extend prior work on combining graded effects and coeffects~\cite{10.1145/2951913.2951939}.

\item[Free grounding order]
  Following the pattern of \rulename{$\ox$\,i}, our rules ground finitely supported variables left to right.
  This choice is arbitrary: there need only be some order which grounds all variables without circularity.
  This ordering can discovered automatically for Datalog programs~\cite{DBLP:conf/ppdp/ContrastinOR18}.
  Extending this to \fslang\ would free the programmer from thinking about grounding order.

\item[Destructuring $\boldsymbol\with$ pairs]
  Outer joins like \name{union} are most naturally expressed as destructuring their $\with$ pair argument,
  but standard linear/relevance typing rules do not allow this.
  We suspect it may be possible by stealing rules from the logic of bunched implications~\cite{DBLP:journals/bsl/OHearnP99}.

\item[Patterns and finite maps]
  We would like to extend finite map $\lambda$ and application to patterns, such as tuples $\fn{(x,y)} t$ and $t\<(x,y)$.
  Otherwise, for example,
  currying/uncurrying of finite maps is quite tricky.%
  \footnotemark{}
  %
  Ideally, application patterns could mix
  grounding finitely supported variables (as in \rulename{$\fto$\,e}) with looking up expressions (as in ${\fto}\,\textsc{e}_2$), e.g.\ $\J{f\<(x,y)} P {\hyp x A} {\hyp f {A \times A \fto P}} {\hyp y A}$.
  However, the unusual behavior of finite support contexts $\O$ makes specifying and implementing finite support patterns challenging.

  \footnotetext{
    The confident reader may take this as a brainteaser.
    Make sure your solution has a typing derivation!
    Our solution (upside-down):

    \noindent
    \rotatebox[origin=c]{180}{
      \parbox{\dimexpr\textwidth - 12pt\relax}{
        The secret is to combine left-to-right grounding with an immediately applied finite $\lambda$:
        \begin{code}
    \name{curry} : (A \times B \fto P) \lto (A \fto B \fto P)\\
    \name{curry} \<f \<a \<b =
      (\fn x f\<x \when \teq{\pi_1\<x}{a} \tand \teq{\pi_2\<x}{b}) \<(a,b)
      \\
    \name{uncurry} : (A \fto B \fto P) \lto (A \times B \fto P)\\
    \name{uncurry} \<f \<\name{ab} =
    (\fn a \fn b f\<a\<b \when \teq{(a,b)}{\name{ab}})
    \<(\pi_1\<\name{ab}) \<(\pi_2\<\name{ab})
  \end{code}

        \phantom{}
      }
    }
  }

\item[Metatheory]
  \fslang\ lacks metatheory.
  Our semantics is a sketch; we have not proven semantic finite support or point preservation, nor syntactic weakening or substitution.
  The need for ${\fto}\,\textsc{e}_2$ should show this is surprisingly subtle.
  Moreover, because of left-to-right grounding, substitution fails for finitely supported variables:
  $(\fn x t) \< u$ may have a typing derivation when $t \{x \mapsto u\}$ does not (see previous footnote's solution for an example).

\todo{\item commutative monoids, how do you declare them?

\item incrementality
\item really nail down the connection with graded monads \& effect/coeffect systems. is our work a special case of theirs? what about the funky asymmetric $\ox$ rule?
  }
\end{description}











\bibliographystyle{splncs04}
\bibliography{fslang}

\end{document}